\documentclass[12pt,draft]{article}
\begin{document}
\title{Reply to a comment "No robust phases in aerogel..." (cond-mat/0505281).}
\author{I.A.Fomin\\
P. L. Kapitza Institute for Physical Problems, \\
ul. Kosygina 2, 119334 Moscow,Russia}
\date{ }
\maketitle
\begin{abstract}
 The arguments of Volovik are refuted.
\end{abstract}
 \bigskip

\textbf{1}. The estimations made in the Comment are based on the
assumption that ABM order parameter is the only relevant minimum
of the Ginzburg and Landau (GL) free energy and its energy is
smaller then that of other possible minima by the energy of the
order of the full condensation energy. This situation is opposite
to the situation considered in the criticized papers [C1], where
competition of nearly degenerate states is assumed (in what
follows references of the Comment are prefixed by a capital C).
Free energy of bulk (without aerogel) superfluid $^3$He has 18
extrema \cite{march} and the situation assumed in the Comment does
not seem to be very realistic.

For a present discussion relative energies of the states,
corresponding to nonferromagnetic equal-spin pairing phases are of
importance. Among the mentioned extrema  there are four minima
meeting this requirement \cite{mermin}. Two of them - ABM and
axiplanar state are so close in energy that identification of
A-phase as ABM-state has been contested in the literature
\cite{gould}. Axiplanar state unlike ABM contains in its vicinity
 robust states, as it was discussed earlier \cite{fomin1}. These
states are also close in energy to the ABM. For a crude estimation
of a relative difference of energies of competing states (to be
referred as $\gamma$ in what follows) weak coupling values of
$\beta_1,...,\beta_5$ parameters were used. With these values a
relative difference of energies of the robust state and ABM
corresponds to $\gamma\sim 1/20$. Strong coupling corrections to
parameters $\beta$ will change the difference, still $\gamma\sim
1/10$ is a fair estimation. Contribution of fluctuations to energy
has to be compared not with the full condensation energy $F_0$ but
with much smaller value $\gamma F_0$. The regular part of this
contribution, which comes from the gapped modes, is of the order
of $\alpha F_0$ in agreement with and in the notations of the
Comment. A value of parameter $\alpha\sim(\eta^2/\sqrt\tau)$  can
be estimated from the measured width of the specific heat jump
\cite{parpia}. According to this data $\alpha\sim 1$ when
$\tau\sim (1/30)$. Because of the weak dependence on $\tau$
everywhere in the GL region parameter $\alpha\approx 1/5$ is at
least comparable or greater then $\gamma$ and even a regular
contribution of fluctuations can mix-up energies of competing
states in a contrast to the statement of the Comment.

\textbf{2}. The main object of criticism in the Comment is a
contribution of fluctuations of Goldstone modes to the energy.
According to the Comment this contribution is of the order of
$\alpha^2F_0$ thus even smaller then the contribution of the
gapped modes so that the free energy is a regular function of
$\alpha$ and the original free fnergy $F_0({A}_{\mu j}^{(0)})$ is
a good starting point for expansion on a small $\alpha$. This
assertion is in a conflict with the Imry and Ma statement [C5]
that the ordered state can be destroyed by "arbitrarily small
random field". It indicates that new free energy $F(\bar{A}_{\mu
j})$ which includes the contribution of fluctuations has to be a
singular function of $\alpha$ and the argument based on continuity
has to be taken with a great care.

The standard procedure \cite{Larkin} of finding of $F(\bar{A}_{\mu
j})$ is based not on the direct averaging of the original free
energy (or of its parts as it is done in the Comment), but on a
derivation of equation for the average order parameter, which in
the present case has the following form:
$$
\tau\bar A_{\mu j}+\frac{1}{2}\sum_{s=1}^5
\beta_s\bigl[\frac{\partial I_s}{\partial A^*_{\mu j}}+
\frac{1}{2}\bigl( \frac{\partial^3 I_s}{\partial A^*_{\mu
j}\partial A_{\nu n}\partial A_{\beta l}} <a_{\nu n}a_{\beta l}>+
$$
$$ 2\frac{\partial^3 I_s}{\partial A^*_{\mu j}\partial A^*_{\nu
n} \partial A_{\beta l}}<a^*_{\nu n}a_{\beta l}>\bigr)\bigr]=
-<a_{\mu l}\eta_{lj}>.              \eqno(1)
$$
Corresponding free energy, if necessary, has to be constructed so
that it generates the derived equation. The averages of
fluctuations of the order parameter $<a_{\nu n}a_{\beta l}>$ in
"Goldstone" directions are proportional to a diverging integral,
i.e. are singular. It has been checked by a direct substitution
that coefficients in front of the singular averages are not
identical zeros. It means that GL equation contains singular
terms. There is no reason for a cancellation of singular terms in
the expression for free energy as well. It should be remarked
though that  free energy has not been used in the arguments of
Refs. [C1].

Volovik in construction of the free energy followed "physical"
argument, which does not take into account important features of
the problem. In particular, he overlooks a fact that Goldstone
directions depend on the average order parameter. As a result
variation of his free energy will not contain terms which have to
be present in the equation (1).

The singular terms in the Eq. (1) being proportional to the
diverging integral are much greater then the regular terms. That
determines a procedure of its solution. As a first step the
principal terms are set to be equal to zero. This condition
selects a degenerate class of robust order parameters. Remaining
terms in the equation are treated as a perturbation, lifting this
degeneracy. They have to be considered on a class of robust order
parameters. So, robust order parameters are asymptotic solutions
of GL equation in a limit $\gamma\rightarrow 0$, $\alpha\ll 1$ and
the ABM order parameter is not solution of this equation in the
considered limit in a contrast to the statement, made in the
Comment. Energies of two states were not compared directly. A
problem of comparison of different states does not arise here
because in the considered limit a family of robust phases is the
only nontrivial extremum of the free energy.

   Summing up one can say that the declared in the Comment error in
overestimation of fluctuations does not exist. The diverging terms
are present in the GL equation and this is sufficient for
selection of the robust phases.  The robust phases are extrema of
the proper free energy. The situation, considered in the Comment
and the one discussed in Refs. [C1] correspond to different
regions of parameters: $\gamma\sim 1$, $\alpha\ll 1$, so that
$\gamma\gg\alpha$ (Comment), $\gamma\leq\alpha\ll 1$ (Refs. [C1]).
For that reason a criticism presented in the first part of the
Comment has no relevance to the problem discussed in the
criticized papers.

About the situation in the real $^3$He it has to be remarked that
the present knowledge of coefficients $\beta_1, ...,\beta_5$ is
not sufficiently accurate for reliable reconstruction of
"topography" of the GL free energy. Even though the given above
estimations show that situation is favorable for realization of
robust phases the competing situation [C7] can not be ruled out
and it can realize in its range of parameters, for example when
aerogel is very dilute if macroscopic description still applies.

I thank A.F. Andreev and V.I. Marchenko for the stimulating
discussions.

  \end{document}